\documentclass{aa}
\usepackage{epsfig,lscape}
\usepackage{graphics}

\newcommand{\zergscma}{10$^{-15}$\,erg\,s$^{-1}$\,cm$^{-2}$\,\AA$^{-1}$}
\begin{document}

\thesaurus{11(11.01.2;11.19.1;11.09.1 NGC\,5548)}

\title{A spectroscopic and photometric study of short-timescale 
       variability in NGC\,5548
\thanks{Based on observations taken at the German-Spanish Astronomical Centre
 Calar Alto, Mt.\,Hopkins Observatory, Crimean Observatory, SAO, University
of Nebraska Observatory}
}

\author{M.\,Dietrich\inst{1,2}
\and C.F.\,Bender\inst{3}
\and D.J.\,Bergmann\inst{3}
\and T.E.\,Bills\inst{3}
\and N.G.\,Bochkarev\inst{4}
\and A.\,Burenkov\inst{5}
\and C.M.\,Gaskell\inst{3}
\and D.D.\,Gutzmer\inst{3}
\and R.\,Grove\inst{3}
\and M.E.\,Hiller\inst{3}
\and J.P.\,Huchra\inst{6}
\and E.S.\,Klimek\inst{3}
\and C.\,Lund\inst{3}
\and N.\,Merkulova\inst{7,8}
\and S.\,Pebley\inst{3}
\and M.A.\,Poulsen\inst{3}
\and V.I.\,Pronik\inst{7,8}
\and S.G.\,Sergeev\inst{7,8}
\and E.A.\,Sergeeva\inst{7,8}
\and A.I.\,Shapovalova\inst{4}
\and V.V.\,Vlasyuk\inst{5}
\and B.\,Wilkes\inst{6}
}
\institute{
Department of Astronomy, University of Florida, 211 Bryant Space Science 
 Center, Gainesville, FL 32611-2055, USA
\and
Landessternwarte Heidelberg, K\"{o}nigstuhl, D-69117 Heidelberg, Germany
\and
University of Nebraska, Lincoln, Department of Physics and Astronomy, 
 Lincoln, NE 68588-0111, USA
\and
Sternberg Astronomical Institute, Universitetskij Prospect, 13, 119899 Moscow,
 Russia
\and
Special Astrophysical Observatory, Russian Academy of Science, Nyzknij Arkhyz,
 Karachaj-Cherkess Republic, 369167, Russia
\and
Harvard-Smithsonian Center for Astrophysics, 60 Garden Street, Cambridge, MA 
02138, USA
\and
Crimean Astrophysical Observatory, p/o Nauchny, 98409 Crimea, Ukraine
\and
Isaac Newton Institute of Chile, Crimean Branch, Chile
}

\offprints{M.Dietrich}
\mail{dietrich@astro.ufl.edu}

\date{Received 17.11.2000 / Accepted 06.03.2001}

\titlerunning{short-term variability of NGC\,5548}
\authorrunning{Dietrich et al.}

\maketitle

\begin{abstract}
Results of a ground-based optical monitoring campaign on NGC\,5548 in June 
1998 are presented. 
The broad-band fluxes ($U$,$B$,$V$), and the spectrophotometric optical 
continuum flux $F_\lambda$(5100\,\AA ) monotonically decreased in flux 
while the broad-band $R$ and $I$ fluxes and the integrated emission-line 
fluxes of H$\alpha$ and H$\beta$ remained constant to within 5\% . 
On June 22, a short continuum flare was detected in the broad band fluxes.
It had an amplitude of about $\sim$18\%\ and it lasted only $\approx$90\,min.
The broad band fluxes and the optical continuum $F_\lambda$(5100\,\AA )
appear to vary simultaneously with the EUV variations.
No reliable delay was detected for the broad optical emission lines in 
response to the EUVE variations.
Narrow H$\beta$ emission features predicted as a signature of an accretion 
disk were not detected during this campaign.
However, there is marginal evidence for a faint feature at 
$\lambda \simeq$\,4962 \AA\ with FWHM$\simeq$6\AA\ redshifted by 
$\Delta $v$\simeq$1100 km\,s$^{-1}$ with respect to H$\beta _{narrow}$.
\end{abstract}

\section{Introduction}
The ultraviolet and optical continuum and the broad emission line flux of
Seyfert\,1 galaxies are known to be variable on timescales of a few days 
until years. 
Variations on timescales of hours and even less have been observed in X-rays 
(e.g.\,Barr \& Mushotzky \cite{BaMu86}; Leighly et al.\,\cite{Leighlyetal96}; 
George et al.\,\cite{Georgeetal98a},\cite{Georgeetal98b}).\\
To study the physical processes which are responsible for the observed 
spectral energy distribution of an active galactic nuclei (AGN)
multiwavelength monitoring campaigns have proven to be an excellent tool 
(cf.\,Peterson \cite{BMP93}; Netzer \& Peterson \cite{NePe97}
for a review; Marshall et al.\,\cite{Marshalletal97}; 
Nandra et al.\,\cite{Nandraetal98}; Edelson et al.\,\cite{Edelsonetal00}). 
Thus, over the last decade, several large space-based and ground-based 
monitoring programs have been undertaken for nearby AGN, such as
NGC\,5548 (Clavel et al.\,\cite{Claveletal91}; 
           Peterson et al.\,\cite{BMPetal91}, \cite{BMPetal92}, 
           \cite{BMPetal94}, \cite{BMPetal99};
           Maoz et al.\,\cite{Maozetal93}; Dietrich et al.\,\cite{MDetal93}; 
           Korista et al.\,\cite{Koristaetal95};
           Chiang et al.\,\cite{Chiangetal00}),
NGC\,3783 (Reichert et al.\,\cite{Reichertetal94}; 
           Stirpe et al.\,\cite{Stirpeetal94}; 
           Alloin et al.\,\cite{Alloinetal95}),
NGC\,4151 (Crenshaw et al.\,\cite{Chrenshawetal96}; 
           Kaspi et al.\,\cite{Kaspietal96}; 
           Warwick et al.\,\cite{Warwicketal96};
           Edelson et al.\,\cite{Edelsonetal96}), 
Fairall\,9 (Rodr\'{\i}guez-Pascual et al.\,\cite{RoPaetal97};
            Santos-Lle\'{o} et al.\,\cite{Santosetal97}),
3C\,390.3 (Leighly et al.\,\cite{Leighlyetal97};
           Dietrich et al.\,\cite{MDetal98}; 
           O'Brien et al.\,\cite{OBrien98}),
and NGC\,7469 (Wanders et al.\,\cite{Ignaz97}; 
               Collier et al.\,\cite{Collieretal98}).

A broad-line region (BLR) size of the order of less than a few light weeks
is indicated for Seyfert\,1 galaxies by the correlated variations of the 
broad emission-line flux and of the optical/ultraviolet continuum.
It is generally assumed that the central supermassive black hole is 
surrounded by an accretion disk.
Such an accretion disk is the probable origin of a significant 
fraction of the broad emission-line flux (e.g.\,Laor \& Netzer \cite{LaNe89}; 
Dumont \& Collin-Souffrin \cite{DuCoSo90}; Halpern \cite{Halpern90}; 
Zheng, Veilleux, \& Grandi \cite{ZhVeGr91}; 
Hubeny et al.\,\cite{Hubenyetal00}).
Recently, Kaspi et al.\,(\cite{Kaspietal00}) published results of a long 
term monitoring campaign of quasars. 
Including the results of Seyfert\,1 galaxies (Wandel et al.\,\cite{Wandel99}) 
they suggested a relation between the BLR size and the optical luminosity 
given by r$\sim$L$^{0.7}$.

The Seyfert\,1 galaxy NGC\,5548 has been continuously monitored in the 
optical since late 1988 by the international AGN watch consortium 
(Peterson et al.\,\cite{BMPetal99}; cf. Alloin (\cite{Alloin94}) for an 
AGN watch overview). 
In June 1998 this prominent Seyfert\,1 galaxy was targeted for a 
coordinated intense monitoring campaign using EUVE, ASCA, and RXTE to study
the high energy continuum emission and its temporal characteristics 
(Chiang et al.\,\cite{Chiangetal00}). 
Chiang et al.'s observations indicate that the variations at $\sim$0.2\,keV 
(EUV) appear to lead similar variations at energies larger than 
$\sim$1\,keV by 3\,--\,8 hrs. This was unexpected as it was generally assumed 
that correlated variations of the EUV, UV, and optical emission would all be 
due to reprocessed higher energy radiation.\\
Since this campaign provided the rare opportunity to access the high energy 
continuum variations  with especially high temporal sampling, we organized a 
simultaneous ground-based campaign for the optical wavelength domain.

Assuming part of the BLR flux is emitted from the accretion disk,
variations on timescales of one day or even less should be detected
(Stella \cite{Stella90}). For this model of a relativistic Keplerian disk 
it was shown that weak narrow features should drift across the emission line 
profile.  
These weak structures are expected to start at high velocities i.e.\, in the 
outer profile wings, and move towards the line center to be due to the
longer time delay at larger radii of the disk in response to the variable 
continuum emission. The timescale of the shift depends on the mass of
the central black hole and it is for NGC\,5548 of the order of less than
several days (cf.\,Stella \cite{Stella90}).
The detection of such structures would be a strong indication for the presence
of an accretion disk in the innermost region of an AGN.

Generally, current monitoring campaigns have not provided the necessary 
temporal and spectral resolution for detecting such substructures in the 
broad emission line profiles. 
However short term monitoring campaigns had been undertaken for NGC\,4151 
(Xanthopoulos \& DeRobertis \cite{XaDeRo91}; 
Crenshaw et al.\,\cite{Chrenshawetal96}; Kaspi et al.\,\cite{Kaspietal96}; 
Warwick et al.\,\cite{Warwicketal96}; Edelson et al.\,\cite{Edelsonetal96}) 
and for a small sample of AGN (e.g.\,NGC\,5548, NGC\,4151, 3C\,390.3, Arp102B,
Mkn\,6; Eracleous \& Halpern \cite{ErHa93}). As yet no significant short 
timescale broad emission line flux variation or profile variability has 
been detected.

In this paper, we present the results of the optical photometric and 
spectroscopic observations that were obtained in June 1998 simultaneous with 
the high energy campaign (Chiang et al.\,\cite{Chiangetal00}).
In section 2 we describe the optical observations and outline intercalibration 
procedures by which a homogeneous set of photometric and spectroscopic 
measurements can be achieved. 
In section 3 we present measurements of the broad-band flux as well as of the 
broad Balmer emission-line flux. 
We compare our results with the results of the simultaneous campaign of the 
EUV wavelength range (Chiang et al.\,\cite{Chiangetal00}).
We also discuss the shape of the H$\alpha $ and H$\beta $ line profiles.
We summarize our results in section 4.

\section{Observations and Data Analysis}
Spectroscopic and broad-band photometric measurements of NGC\,5548 were 
obtained by several groups in June 1998.
Table 1 gives a brief overview of the individual participants who recorded 
the data we report here. 
Each group (column 1) was assigned an identification code given in 
column (2) which is used throughout this paper. This code was based on
that used for the ongoing NGC\,5548 monitoring campaign (Peterson et 
al.\,\cite{BMPetal99}). Column (3) gives the aperture of the telescope used. 
Columns (4) -- (8) list the sizes of the focal-plane apertures used in various
observations; fixed instrument apertures were used.
In column (9), the spectrograph slit width (in the dispersion direction) and 
extraction width (cross-dispersion dimension) of the spectra are listed.

\begin{table*}
\caption[]{Overview of observations; the aperture size is given in units 
           of arcsec}
\begin{tabular}{lcccccccc}
\hline
Source&Code&Tel.&\multicolumn{5}{c}{Photometry Aperture}&Spectroscopy\\
      &    &[m] & U& B& V& R& I& Aperture\\
\hline
Mt.\ Hopkins Observatory     &C & 1.5&---&---&---&---&---&   3x4.6 \\
Crimean Astrophysical Obs.   &D &1.25&15 & 15& 15&15 &15 & --- \\
Calar Alto Observatory       &G & 2.2&---&---&---&---&---&2.06x2.61\\
Special Astrophysical Obs.   &L1& 1.0&---&---&---&---&---&8x6.6,8x19.8\\
Special Astrophysical Obs.   &L2& 6.0&---&---&---&---&---&2x6      \\
Univ.\, of Nebraska 0.4\,m   &Q & 0.4&---&---& 8 &---&---& --- \\
Shajn reflector, Crimean Obs.&W & 2.6&---&---&---&---&---&   3x11  \\
\hline
\end{tabular}
\end{table*}

A complete log of the photometric and spectroscopic observations is given
in Table 1A and 2A (available in electronic form at the CDS).

\subsection{Optical Photometry}
Photometric observations of NGC\,5548 were obtained by Gaskell et al. 
(sample Q) and Merkulova (sample D). 
The brightness of NGC\,5548 was determined with respect to stars in 
the field of NGC\,5548 using the photometric sequence (stars 1 and 2) 
defined by Penston, Penston, \& Sandage (\cite{PePeSa71}), and star C1 and C 
defined by Lyuty (\cite{Lyuty72}). The bright star $\approx$1\arcmin\ east 
to the galaxy located at P.A.\,=\,240$^{\degr}$ at a distance of 
$\approx$9\arcmin\ with respect to NGC\,5548 also was used. This star is 
referred in the HST guidestar catalogue as gsc\,0201001062 
(RA = 17$^h$\,17$^m$\,29\fs 66, Dec = $+$25\degr 03\arcmin 10\farcs 5 
(2000.0)). 
Star C1 in Lyuty is identical to star 4 in Penston, Penston, \& Sandage. 
The U,\,B,\,V magnitudes of C1 and 4 are identical within the errors and 
differ by only $\pm$0.02 mag (B,V) and $\pm$0.17 mag (U)
(cf.\,Tab.\,2). Star C is the bright star close to star 3 in Penston, 
Penston, \& Sandage located approximately 1\farcm 3 southeast of it.

\begin{table}
\caption[]{Adopted magnitudes for standard stars}
\begin{tabular}{cccccc}
\hline
\noalign{\smallskip}
band&  1$^a$  &  2$^a$  & 4,C1$^b$&  C$^b$  &  0  \\
\noalign{\smallskip}
\hline
 U  &14.52&16.24&13.23&11.17& --- \\
 B  &14.45&16.07&12.78&10.96& --- \\
 V  &13.80&15.38&11.90&10.47&11.29\\
 R  & --- & --- &11.17& 9.96& --- \\
 I  & --- & --- &10.64& 9.57& --- \\
\noalign{\smallskip}
\hline
\end{tabular}

\medskip
\small{$^a$from Penston, Penston, \& Sandage (\cite{PePeSa71}), estimated 
       uncertainty $0.02$\,mag.}\\
\small{$^b$from Lyuty (\cite{Lyuty72}), estimated uncertainty $0.01$\,mag. 
       ({\it U}\,) of C1 $0.02$\,mag.}
\end{table}

The photometric observations of NGC\,5548 of sample D obtained with the 
1.25\,m telescope of the Crimean Astrophysical Observatory were recorded 
using a five channel version of the Double Image Chopping Photometer - 
Polarimeter (Piirola \cite{Piirola73}).
The observations were made simultaneously in five colors using dichroic
filters to split the light into five spectral bands.
The resulting passbands are close to the standard Johnson UBVRI photometric 
system with effective wavelengths at 3600\,\AA , 4400\,\AA , 5300\,\AA, 
6900\,\AA , and 8300\,\AA , respectively. A diaphragm with two equal 
apertures (15\arcsec \ in diameter) was used in the focal plane of the 
1.25\,m telescope.
The distance between the aperture centers was 26\arcsec . A rotating chopper
alternately closes one of the apertures, leaving the other free, thus the 
photocathode is illuminated alternately by the galaxy (or the star) and the 
sky apertures. Because the centers of the apertures are closely spaced, 
background observations were also obtained  at a distance of about 7\arcmin \ 
from the galaxy nucleus to correct the observational background for a 
contribution of the outer regions of the galaxy. The telescope is fully 
automated and an autoguider was used. 
Observations of the nucleus of NGC\,5548 were made on nights with good 
atmospheric conditions --- when the estimated seeing was in the range of  
1\arcsec \ to 3\arcsec .
The positional accuracy by the autoguider during observations is better than 
20\%\ of the estimated seeing, i.e.\, better than 0\farcs 6 in the worst 
case.
The galaxy nucleus was also positioned in the aperture using the autoguider, 
with similar positioning errors.
The conventional technique of differential measurements was applied.
Two comparison stars (labelled as C1 and C) were taken from the list of 
Lyuty (\cite{Lyuty72}). Observations were performed against a comparison star 
C1; and the second star, C, was used to check the results of the first 
comparison.
In addition to stars C1 and C, secondary $UBVRI$ standards by Neckel \& 
Chini (\cite{NeCh80}) were used for an absolute calibration.
The measurements were carried out by observing in the following sequence,
C--C$_{1}$--sky--AGN--sky--C$_{1}$--C--sky.
The time resolution was about 3.5 minutes. During a single observation, 8 
integrations of 10 sec each were made.  Photon statistics (corrected for sky 
background) were applied to calculate photometric errors, which generally were
the same as the rms errors obtained by averaging the 8 integrations.
Generally, the atmospheric seeing during the observations obtained with the 
1.25\,m CAO telescope averaged 2\arcsec . It was better for two nights 
(June 27/28 and June 30/July 1) and worse for the night June 22/23 
(2-3 \arcsec ). 

The V-band observations of Gaskell et al.\,(sample Q) were obtained with the 
University of Nebraska 0.4\,m telescope in Lincoln, Nebraska. 
The frames were recorded with a ST-7 CCD camera. 
The flux of NGC\,5548 was determined for an 8\arcsec\ diameter aperture. 
The sky was measured in an annulus of radius 15 to 20 \arcsec .
The V-band magnitude was derived relative to the comparison stars 1 and 
gsc0201001062 in the field of NGC\,5548 (cf.\,Penston, Penston, \& Sandage 
\cite{PePeSa71}). 
The exposure time of an individual frame was 3 minutes. 
Generally, to increase the signal-to-noise ratio of the flux measurements, 
the V-band magnitudes of three subsequently recorded frames were averaged.
The seeing during the observations was typically 4\arcsec . There was no
correlation detected between the derived magnitude of NGC\,5548 and the 
seeing.

To combine the broad band flux measurements of sample D and Q, the apparent 
magnitudes were transformed into fluxes. The conversion was performed using 
the following equations (Allen \cite{Allen73}; Wamsteker \cite{Wamsteker81}):
\begin{eqnarray}
\log F_\lambda (3600\,{\rm \AA}) & = & -0.4 m_U - 8.361,   \\
\log F_\lambda (4400\,{\rm \AA}) & = & -0.4 m_B - 8.180, \nonumber \\
\log F_\lambda (5500\,{\rm \AA}) & = & -0.4 m_V - 8.439, \nonumber \\
\log F_\lambda (7000\,{\rm \AA}) & = & -0.4 m_R - 8.759, \nonumber \\
\log F_\lambda (9000\,{\rm \AA}) & = & -0.4 m_I - 9.080. \nonumber
\end{eqnarray}

The broad band V-fluxes of sample Q were scaled to the V-band fluxes 
observed for sample D. The additive scaling factor was determind by 
comparison of observations which were separated by less than 2 hours.
We derived a scaling factor of 
(2.944$\pm$0.524)\,$10^{-15}$\,erg\ s$^{-1}$\,cm$^{-2}$\,\AA$^{-1}$
based on 5 epochs which corresponds to $\Delta $mag\,=\,0.20$\pm$0.02. 
The difference is caused by the significantly different sizes of the
apertures used to measure the flux of NGC\,5548 (Tab.\,1).
The resulting  $U$,$B$,$V$,$R$,$I$ broad-band light curves are given in 
Table 1A (available in electronic form at the CDS) in units of \zergscma\ 
and are displayed in Figure 1.

\begin{figure}
\resizebox{\hsize}{!}{\includegraphics[bb=33 35 373 780]{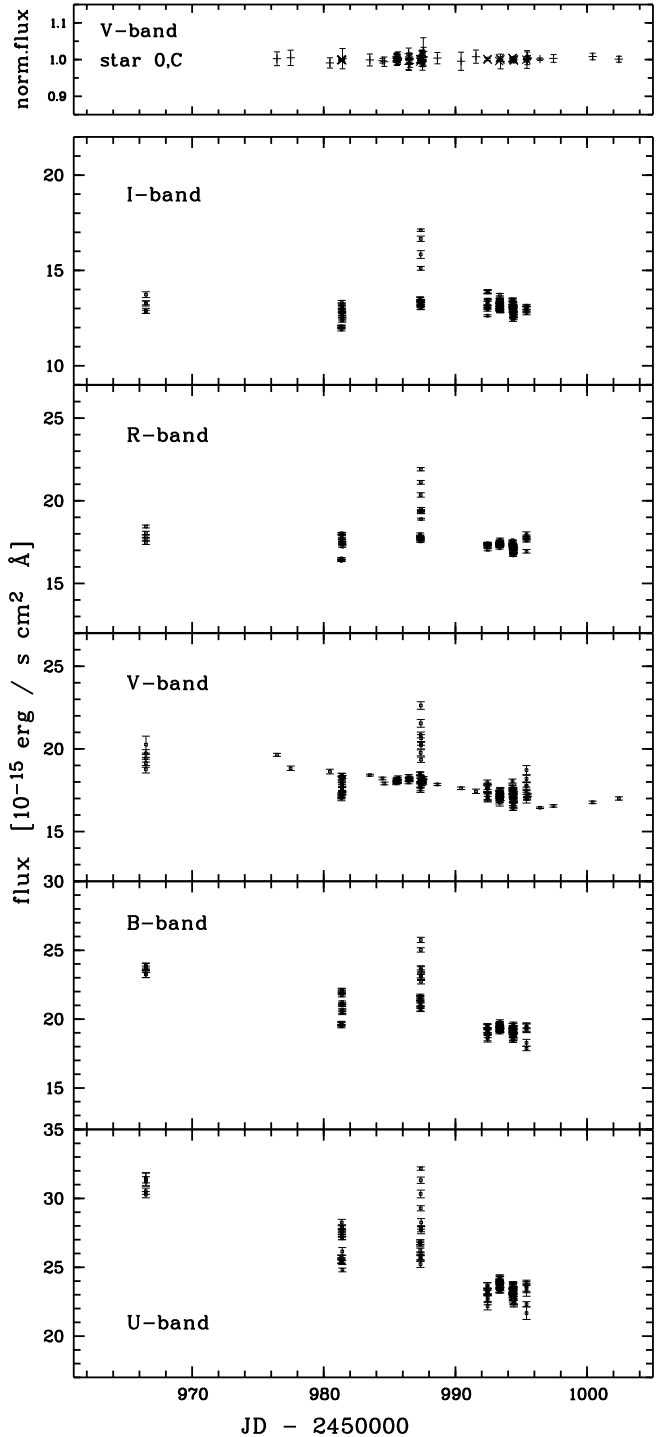}}
\caption{Optical broad-band light curves for NGC\,5548. Fluxes 
in broad-band $U$,\,$B$,\,$V$,\,$R$, and $I$  are in units of 
$10^{-15}$\,erg s$^{-1}$\,cm$^{-2}$\,\AA$^{-1}$.
In the top panel the normalized V-band flux of star 0 (+) and 
C (x) is shown which was contant within $\sim$0.57\,\% .}
\end{figure}

\subsection{Optical Spectroscopy}
The flux calibration of AGN spectra can be accomplished in several ways.
In variability studies, it has become common practice to normalize the flux 
scale to the fluxes of strong forbidden narrow emission lines which are 
assumed to be constant over timescales of at least several decades 
(e.g.\,Peterson \cite{BMP93}). 
This assumption is justified by the large spatial extent and low gas density 
of the narrow-line region (NLR), but is only valid if the aperture used
is larger than the NLR.
Light travel-time effects and the long recombination timescale 
($\tau_{\rm rec} \approx 100$\,years for $n_e \approx 10^3$\,cm$^{-3}$) 
damp out short timescale variability. 

As a major sources of uncertainty in inhomogenous samples of variability data
is the use of different instrumental settings, it is important to take 
aperture effects into account (Peterson \& Collins \cite{PeCo83}). 
The seeing-dependent uncertainties which are introduced by the aperture 
geometry can be minimized by using large apertures. 
It has been shown that apertures of 5\arcsec\ $\times$ 7\farcs 5 can 
reduce seeing-dependent photometric errors to no more than a few percent 
in the case of nearby AGNs (Peterson et al.\,\cite{BMPetal95}).

The spectra of NGC\,5548 obtained at the 1\,m telescope of SAO RAS 
(sample L1 Table 1) were extracted for a slit aperture of 
8\farcs 0 $\times$ 6\farcs 6 and 8\farcs 0 $\times$ 19\farcs 8, 
respectively. For these large apertures seeing losses can be neglected. 
To measure the 
F({[O\,{\sc III}]} 5007\,\AA ) a linear pseudo-continuum was fitted beneath 
the {[O\,{\sc III}]} emission line and the flux was measured for the 
wavelength range 
$\lambda \lambda$5062-5108 \AA . For the extraction windows used for sample 
L1 we determined a flux of F({[O\,{\sc III}]} 5007\,\AA ) 
= (5.07$\pm$0.05) 10$^{-13}$erg\,s$^{-1}$\,cm$^{-2}$ 
(8\farcs 0 $\times$ 6\farcs 6) and  
F({[O\,{\sc III}]} 5007\,\AA )
= (5.91$\pm$0.05) 10$^{-13}$erg\,s$^{-1}$\,cm$^{-2}$ 
(8\farcs 0 $\times$ 19\farcs 8), respectively. These values are 
in good agreement with F({[O\,{\sc III}]} 5007\,\AA ) measured for NGC\,5548
during the last 10 years (cf.\,Peterson et al.\,\cite{BMPetal99}). For large 
apertures of comparable size (e.g.\, 8\farcs 8 $\times$ 12\farcs 8)  
F({[O\,{\sc III}]} 5007\,\AA )
= 5.1 to 5.9 10$^{-13}$erg\,s$^{-1}$\,cm$^{-2}$ is mentioned
(Peterson et al.\,\cite{BMPetal92}).
To be constistent with the H$\beta$ line flux measurements of NGC\,5548 
obtained since 1988 (Peterson et al.\,\cite{BMPetal99}) we adopted 
F({[O\,{\sc III}]} 5007\,\AA )
= (5.58$\pm$0.27) 10$^{-13}$erg\,s$^{-1}$\,cm$^{-2}$.

\subsection{Intercalibration of the Spectra}
Since the spectra of the various samples were taken with different instruments 
in different configurations, it was necessary to intercalibrate them to a 
common flux level. 
As we have shown above, the {[O\,{\sc III}]}\,5007 line flux was constant to 
better than 1\%\ for large extraction windows. We can safely use the narrow 
emission lines such as the flux of {[O\,{\sc III}]}\,5007 as standards. 
In order to avoid any wavelength-dependent calibration errors, each spectrum 
was scaled in flux locally over a limited wavelength range prior to 
measurement.
The H$\beta$ spectral region was scaled with respect to the 
[O\,{\sc III}]\,4959,5007 line fluxes. 
The H$\alpha$ region was scaled with respect to the flux of the 
[O\,{\sc I}]\,6300, 
[N\,{\sc II}]\,6548,6583, and [S\,{\sc II}]\,6716,6731 emission lines. 
The spectra were intercalibrated using the method described by van 
Groningen \& Wanders (\cite{ErnstIgnaz92}). This procedure corrects for 
different flux scales, small wavelength shifts, and different spectral 
resolutions by minimizing the narrow-line residuals in difference spectra 
formed by subtracting a ``reference spectrum'' from each of the observed 
spectra.
The rescaled spectra were used to derive integrated emission-line fluxes. 

\begin{table}
\caption[]{Scaling factors for photometric and spectroscopic subsets}
\begin{tabular}{lccc}
\hline
\noalign{\smallskip}
sample&add.\, constant&point source&extended source\\
      &V-mag.$^a$ &correction $\varphi ^b$&correction G$^b$\\
\noalign{\smallskip}
\hline
C & ---  &0.972$\pm$0.022&2.02$\pm$0.33 \\
D & 1.000& ---           & ---          \\
G & ---  &0.865$\pm$0.022&3.36$\pm$0.54 \\
L1& ---  &1.00           &0.00          \\
L2& ---  &0.945$\pm$0.025&2.60$\pm$0.53 \\
Q &2.944$\pm$0.524& ---  &---           \\
W & ---  &0.950$\pm$0.028&1.52$\pm$0.32 \\
\noalign{\smallskip}
\hline
\end{tabular}

\medskip
$^a$relative to sample D\\
$^b$relative to sample L1
\end{table}

\section{Results}
\subsection{Light Curves}
The broad band flux variations in $U$,$B$,$V$,$R$,$I$ are displayed in Fig.\,1.
In addition to the broad band flux variations the light curve of the
measured flux of the comparison stars C and 0 (cf.\,Tab.\,2) is shown. 
The normalized flux remained constant within $\sim$0.57\,\% .
The broad band flux of NGC\,5548  decreases throughout the campaign in June 
1998.
This trend is clearly visible in the U- and B-band, and the ratio of the
flux levels observed in early and late June becomes smaller for 
increasing wavelength. In the I-band the flux decay is negligible.
The merged V-band light curve of sample D and Q shows different internal
scatter of the measurements recorded during individual nights (Figs.\,1,2).
This difference might be caused by different seeing conditions of the
observations. Merkulova (sample D) measured the brightness of NGC\,5548
with an 15\arcsec\ aperture under a typical seeing of 1-3\arcsec . The
seeing amounts to $\approx$4\arcsec\ for the measurements Gaskell
et al.\,(sample Q) observed for an 8\arcsec\ aperture. 
Modelling the host galaxy of NGC\,5548 Romanishin et al.\,(\cite{Romanetal95})
showed that the light loss of the AGN is of the order of 10\%\ for sample Q 
and only slightly larger for the host galaxy.
However, the different scatter may be also intrinsic. The temporal 
resolution of the samples amounts to $\leq$4 minutes (D) and nearly 
16 minutes (Q), respectively. 

\begin{figure}
\vspace*{5mm}
\resizebox{\hsize}{!}{\includegraphics[bb=28 246 328 780]{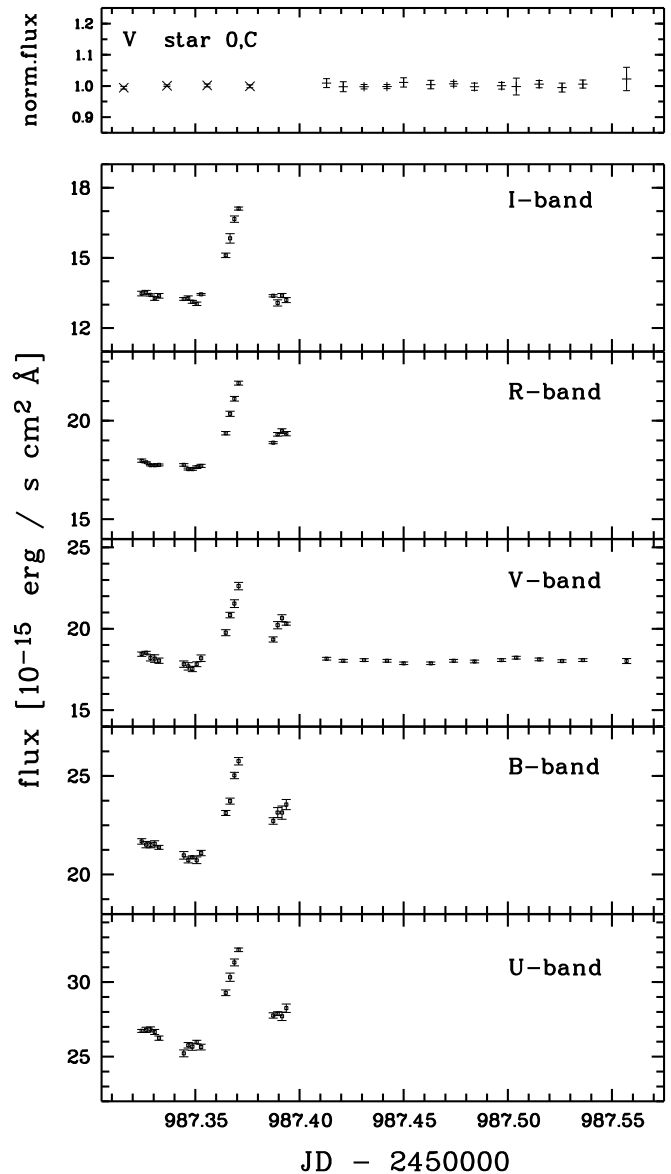}}
\caption{The strong variation in the broad band fluxes as measured on June 22 
         (JD=2450987). In all 5 broad band fluxes a rapid increase and 
         subsequent decay was detected within $\approx$90 minutes.
         In the top panel the normalized V-band flux of star 0 (+) and 
         C (x) is shown.}
\end{figure}

\noindent
The scatter of the brightness during
individual nights is $\approx$5\% for sample D. This amplitude
is of the order which is expected for microvariability. For several
Seyfert\,1 galaxies (NGC\,3516, NGC\,4151, NGC\,5548, NGC\,7469) 
detection of microvariability was claimed on timescales of less than 
10 minutes (cf.\,Lyuty et al.\,\cite{Lyutyetal89}; 
Dultzin-Hacyan et al.\,\cite{Dultzinetal92};
Jang \& Miller \cite{JaMi97}; Noble et al.\,\cite{Nobleetal97}; 
Carini et al.\,\cite{Carinietal98}; 
Welsh et al.\,\cite{Welshetal98}; Ghosh et al.\,\cite{Ghoshetal00}; 
Merkulova \cite{Nelly00}) and amplitudes of the order of 15\%\ to 20\% . 
However, it seems that onset and disappearance of microvariability
follow a random process since several attempts to detect or to confirm 
rapid optical variations of Seyfert\,1 galaxies failed 
(cf.\,Dultzin-Hacyan et al.\,\cite{Dultzinetal93}; 
Petrucci et al.\,\cite{Petruccietal99}). Hence, it might 
be that rapid variations can not be detected on timescales larger than 
15 minutes.

The most obvious structure in the broad band light curves is a flare
like event. This variation was detected on JD\,=\,2450987 (June 22). The 
observations recorded during this night are shown in more detail in Fig.\,2. 
In all 5 broad band measurements a short intense increase of the flux level 
occured. The broad band flux increased by 20$\pm$2\% within $\approx$ 30 
minutes and decreased within $\approx$60 minutes to the flux level of the
beginning of this event. The duration of this flare lasted only 
$\approx$90 minutes. This remarkable flux increase can not be due
to temporal variations of a standard star which was used for relative flux 
calibration. The flux calibration is based on two stars which were in cross 
checked with secondary standard stars taken from Neckel \& Chini 
(\cite{NeCh80}) in addition. 
As can be seen in Fig.\,1 and 2 the used comparison stars were constant 
     within less than 0.57\,\% .
The standard deviations of the primary star used for calibration during 
this night amount to 0.018,\,0.014,\,0.009,\,0.023,\,0.001 mag in 
U,\,B,\,V,\,R,\,I.

Broad emission-line fluxes were integrated over a wavelength range of 
$\lambda \lambda $4850--5020 \AA\ for H$\beta $ and 
$\lambda \lambda $6500--6800 \AA\ for H$\alpha $.
A local linear continuum fit was interpolated under each emission line. In 
the case of the H$\beta$ region, the continuum was defined by the flux measured
in two windows (10\AA\ width) at 4845\,\AA\ and 5170\,\AA\ in 
the observed frame (H$\beta $) and at 6350\,\AA\ and 6965\,\AA\ (H$\alpha $).
No attempt was made to correct any of the measured emission-line fluxes 
for their respective narrow-line contributions.
The optical continuum flux F$_\lambda $(5100\,\AA) was determined as the 
average value in the range 5185--5195\,\AA\ (cf.\,Tab.\,4).

\begin{table}
\caption[]{Integration limits}
\begin{tabular}{lc}
\hline
\noalign{\smallskip}
feature&wavelength range\\
       &  [\AA ]\\
\noalign{\smallskip}
\hline
H$\alpha\,\lambda6563$	& 6500--6800 \\
H$\beta\,\lambda4861$   & 4850--5020 \\
F$_\lambda$(5100\,\AA)  & 5185--5195 \\
\noalign{\smallskip}
\hline
\end{tabular}
\end{table}

The measured H$\beta$ and H$\alpha$ emission line fluxes were adjusted to take
into account the different slit sizes (Tab.\,1).
The measurements of both H$\beta$ and H$\alpha$ were normalized by a similar
factor\\

\begin{equation}
F_{H\beta} = \varphi \, F_{[O\,{\sc III}]} 
\left[ \frac{F_{H\beta}}
{F_{[O\,{\sc III}]}} \right]_{\rm obs}
\end{equation}

The continuum fluxes F$_\lambda$(5100\,\AA) were then adjusted for different 
amounts of host-galaxy contamination (see Peterson et al.\,\cite{BMPetal95} 
for a detailed discussion) by

\begin{figure}
\vspace*{5mm}
\resizebox{\hsize}{!}{\includegraphics{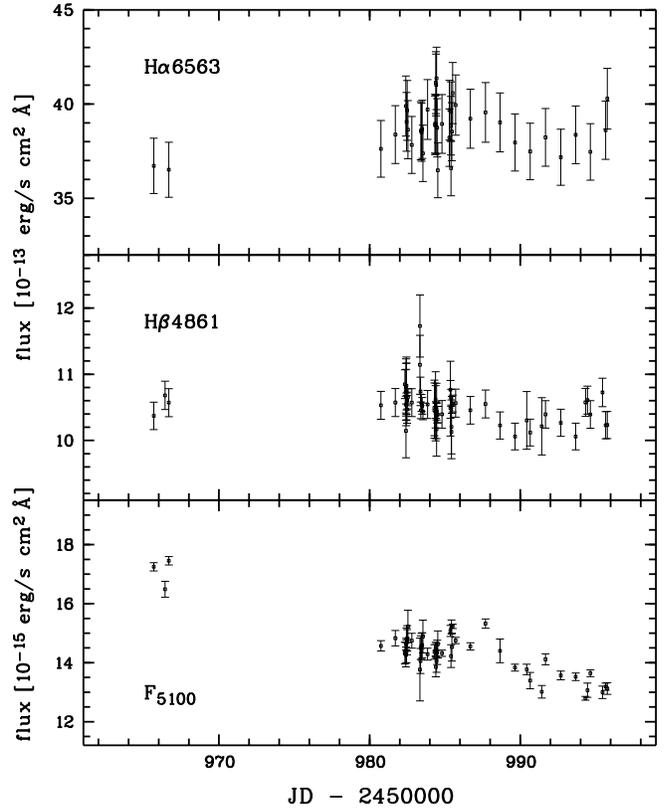}}
\caption{Light curves for the emission lines H$\alpha$, H$\beta$, and
the optical continuum flux $F_{\lambda}$(5100\,\AA). The vertical scale is 
in units of $10^{-13}$\,erg\ s$^{-1}$\,cm$^{-2}$\,\AA$^{-1}$ for the lines 
and $10^{-15}$\,erg\ s$^{-1}$\,cm$^{-2}$\,\AA$^{-1}$ for the continuum.}
\end{figure}

\begin{equation}
F_{\lambda}(5100\,{\rm \AA}) = \varphi \,F_{[O\,{\sc III}]} 
\left[ \frac{F_{\lambda}(5100\,{\rm \AA})}
{F_{[O\,{\sc III}]}} \right]_{\rm obs} + G,
\end{equation}

\noindent 
where $F_{[O\,{\sc III}]}$ is the adopted absolute flux of 
F({[O\,{\sc III}]} 5007\,\AA ), the quantity in brackets is the observed 
continuum to F({[O\,{\sc III}]} 5007\,\AA ) flux ratio measured from the 
spectrum, and $G$ is an aperture-dependent correction for the host-galaxy flux.
The sample L1 which uses a relatively large aperture
(8\farcs 0$\times$ 6\farcs 6) was adopted as the standard 
(i.e., $\varphi$=1.0, $G = 0$ by definition), and other data sets were merged
progressively by comparing measurements based on observations made during
the same night.
Note that this means that any real variability that occurs on timescales this
short will be somewhat suppressed by the process that allows us to combine the 
different data sets. 
The scaling factor $\varphi$ and the additive scaling factor $G$ for the 
various samples are given in Table 3.

The resulting light curves of the optical continuum F$_\lambda$(5100\,\AA) and
of the H$\beta$ and H$\alpha$ emission line fluxes are displayed in Fig.\,3 
and presented in Table 3A (available in electronic form at the CDS). The 
optical continuum light curve 
is very similar to the broad-band V light curve (cf.\,Figs.\,1,3). 
The difference in the flux levels is probably due to the larger aperture used 
for the broad band flux determination relative to that used for the optical 
continuum fluxes. 

A final check of the uncertainty estimates was performed by examining the 
ratios of all pairs of photometric and spectroscopic observations which were 
separated by 0.05\,days or less.\\
There are more than a dozen independent pairs of measurements separated by 
less than 0.05\,d.
For H$\alpha$, the uncertainty estimate is dominated by the spectra taken at
Calar Alto Observatory, while for H$\beta $ more than 20 pairs could be used.
The dispersion about the mean (unity), divided by $\sqrt 2$, provides an 
estimate of the typical uncertainty in a single measurement 
($\sigma _{\rm est}$).
The observational uncertainties ($\sigma _{\rm obs}$) assigned to the spectral
line flux measurements were estimated from the error spectra which were 
calculated by the intercalibration routine, as well as from the 
signal-to-noise ratio within the spectral range near the individual emission 
lines. 
For example the mean fractional flux error is $\sigma _{\rm obs}$=0.040 for 
the H$\alpha $ line, and the average fractional uncertainty from the internal 
statistical estimate is $\sigma _{\rm est}$ = 0.025. 
For the H$\beta $ line the values are $\sigma _{\rm obs}$=0.022 and 
$\sigma _{\rm est}$ = 0.010 which implies that the error estimates for both 
lines are probably quite good.
Generally, the estimated errors ($\sigma_{\rm est}$) are of the same order as,
but slightly smaller than, the observational uncertainties 
($\sigma_{\rm obs}$) derived directly from multiple measurements (Tab.\,5). 
The large difference of $\sigma_{\rm obs}$ and $\sigma_{\rm est}$ for
the broad band photometric fluxes is mainly due to the measurements
obtained on JD=2450987 (June 22), which we expect is due to a real
variation.

\begin{table}
\caption[]{Comparison of uncertainty estimates}
\begin{tabular}{lcclcc}
\hline
\noalign{\smallskip}
feature&$\sigma_{\rm obs}^a$&$\sigma_{\rm est}^b$&feature&$\sigma_{\rm obs}^a$&
 $\sigma_{\rm est}^b$\\
\noalign{\smallskip}
\hline
$U$ &$0.008$&$0.021$&$I$       &$0.006$&$0.024$ \\
$B$ &$0.008$&$0.020$&F$_\lambda$(5100\,\AA)&$0.020$&$0.016$ \\
$V$ &$0.010$&$0.020$&H$\beta$  &$0.022$&$0.010$ \\
$R$ &$0.004$&$0.016$&H$\alpha$ &$0.040$&$0.025$ \\
\noalign{\smallskip}
\hline
\end{tabular}

\medskip
$^a$observational uncertainty based on uncertainties assigned to individual 
    points.\\
$^b$mean fractional uncertainty based on point-to-point differences between 
    closely spaced (i.e., $\Delta t \leq 0.05$\,d) measurements.
\end{table}

The average interval between measurements of the combined broad band light
curves is about $0.24\pm 1.53$\,days for $U$, $B$, $R$, $I$, and for $V$ 
$0.20\pm 0.87$\,days.
However, if the large gaps ($>$5\,days) are not taken into account, the 
sampling rate drops to $0.03\pm 0.15$\,days ($U$,$B$,$R$,$I$) and
$0.15\pm 0.47$\,days ($V$) (Tab.\,6).
The sampling of the H$\alpha $, H$\beta $, and the optical continuum flux
F$_\lambda$(5100\,\AA) can be obviouly devided into two regimes. 
Taking into account all measurements the sampling is about 
$\Delta t$(H$\beta )$ = 0.59$\pm 1.96$\,days for H$\beta $ and 
F$_\lambda$(5100\,\AA).
Neglecting the large temporal gap of nearly two weeks at the beginning of the 
monitoring campaign (full moon), $\Delta t$ is 
$\Delta t$(H$\beta )$ = 0.32$\pm 0.36$\,days. 
For the H$\alpha $ light curve the corresponding sampling rates are
$\Delta t$(H$\alpha ) = 0.81\pm 2.28$\,days and 
$\Delta t$(H$\alpha ) = 0.45\pm 0.43$\,days, respectively.

\begin{table}
\caption[]{Sampling characteristics}
\begin{tabular}{lccc}
\hline
\noalign{\smallskip}
feature&N&interval&interval$^a$\\
       & & [days] & [days]     \\
\noalign{\smallskip}
\hline
$U$       &120&$0.24\pm 1.53$&$0.03\pm 0.15$ \\
$B$       &120&$0.24\pm 1.53$&$0.03\pm 0.15$ \\
$V$       &178&$0.20\pm 0.87$&$0.15\pm 0.47$ \\
$R$       &120&$0.24\pm 1.53$&$0.03\pm 0.15$ \\
$I$       &120&$0.24\pm 1.53$&$0.03\pm 0.15$ \\
F$_\lambda$(5100\,\AA)& 52&$0.59\pm 1.96$&$0.32\pm 0.36$ \\
H$\beta$  & 52&$0.59\pm 1.96$&$0.32\pm 0.36$ \\
H$\alpha$ & 38&$0.81\pm 2.28$&$0.45\pm 0.43$ \\
\noalign{\smallskip}
\hline
\end{tabular}

\medskip
$^a$neglecting temporal gaps larger than 5 days
\end{table}

The broad H$\alpha $ and H$\beta $ emission lines, the optical continuum
flux F$_\lambda$(5100\,\AA) and the $U$,$B$,$V$,$R$,$I$ broad band fluxes 
appear to exhibit small amplitude variations on timescales of days  
(Figs.\,1-3).
The broad band light curves $U$,$B$,$V$, as well as the F$_\lambda$(5100\,\AA )
flux show a decreasing flux from the beginning of this campaign to the end.
The broad band flux at longer wavelengths ($R$ and $I$) appeared to be nearly 
constant. 

The H$\alpha $ and H$\beta $ light curves show  weak evidence for small
amplitude variations on timescales of only a few days. However, this
is statistically not significant; within the
errors the H$\alpha $ and H$\beta $ emission line flux can be regarded as 
nearly constant. This can be seen by deriving the variability parameter
F$_{var}$.\\
The variability parameters $F_{\rm var}$ and $R_{\rm max}$ have been
calculated for the broad-band flux variations as well as for the
broad emission lines and the optical continuum 
(cf.\ Clavel et al.\,\cite{Claveletal91};
Rodr\'{\i}guez-Pascual et al.\,\cite{RoPaetal97}). 
The quantity $R_{\rm max}$ is the ratio of the maximum to the minimum 
flux. The quantity $F_{\rm var}$ is an estimation of the fluctuations of the 
intrinsic variations relative to the mean flux. Therefore, the rms of the 
light curves has been corrected with respect to the uncertainties introduced 
by the observations. 
Values of $R_{\rm max}$ and $F_{\rm var}$ for the broad-band measurements 
are given in Table 7. 
The value of F$_{\rm var}$ for the optical continuum F$_\lambda$(5100\,\AA )
agrees well with F$_{\rm var}$ for the V-band.
Furthermore, for the broad band variations and the optical continuum 
we reproduce the well known trend  that the amplitude of the variation 
decreases with increasing wavelength (Tab.\,7).
The broad band fluxes and the F$_\lambda$(5100\,\AA) continuum flux
were not corrected for the host galaxy contribution since this
fraction is only of the order of less than 10\,\%\, 
(Romanishin et al.\, \cite{Romanetal95}).
The values of F$_{var}$ determined for the H$\alpha $ and H$\beta $ 
emission line flux variations indicated that the Balmer line flux can be 
taken as constant within the errors, even if one is tempted to glimpse hints 
of small amplitude variations on timescales of a few days 
(cf.\,Tab.\,7, Fig.\,3).

\begin{table}
\caption[]{Variability statistics of the light curves;
continuum flux and broad band fluxes in units of 
$10^{-15}$\,erg\ s$^{-1}$\,cm$^{-2}$\,\AA$^{-1}$; line fluxes in units of 
$10^{-13}$\,erg\ s$^{-1}$\,cm$^{-2}$. The EUV flux is given in units of
counts per second.
}
\begin{tabular}{lcccc}
\hline
\noalign{\smallskip}
feature   &mean flux&rms flux&R$_{max}$&F$_{var}$\\
\noalign{\smallskip}
\hline
$U$       &$24.80$&$2.37$&$1.48$&$0.0865$\\
$B$       &$20.08$&$1.56$&$1.44$&$0.0697$\\
$V$       &$17.76$&$0.91$&$1.38$&$0.0421$\\
$R$       &$17.54$&$0.78$&$1.34$&$0.0406$\\
$I$       &$13.15$&$0.69$&$1.44$&$0.0557$\\
F$_\lambda$(5100\,\AA)&$14.4 $& $0.9$&$1.36$&$ 0.040$\\
H$\beta$  &$10.5 $& $0.2$&$1.11$&$-0.005$\\
H$\alpha$ &$38.7 $& $1.2$&$1.13$&$-0.006$\\
EUV       &$0.0993$&$0.0376$&$59$&$0.2035$\\
\noalign{\smallskip}
\hline
\end{tabular}
\end{table}

\begin{figure}
\vspace*{10mm}
\hspace*{2mm}
\resizebox{85mm}{!}{\includegraphics{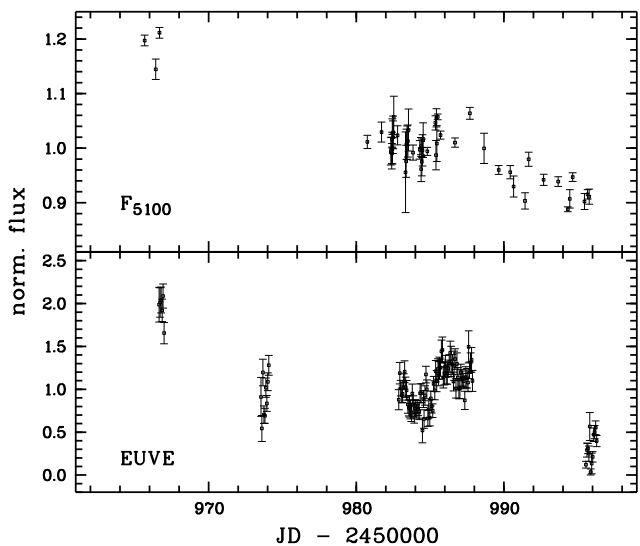}}
\caption{The normalized EUVE light curve from the monitoring program for 
NGC\,5548 (Chiang et al.\,\cite{Chiangetal00}) is shown in the bottom 
panel. In the top panel the normalized variations of F$_\lambda$(5100\,\AA )
are displayed.}
\end{figure}

For comparison we also calculated F$_{var}$ and R$_{max}$ for the rapid 
variations detected by Chiang et al.\,(\cite{Chiangetal00}) in the extrem
ultraviolet domain (Tab.\,7).
In Fig.\,4 we show the normalized light curve of F$_\lambda$(5100 \AA )
and of the EUV variations (kindly provided by Chiang and collaborators).
The F$_\lambda$(5100 \AA ) continuum flux show the same overall trend as 
the EUV variations. But in the optical continuum the rapid variations
visible in the EUV are smeared out or the amplitude of the variations at 
$\lambda$=5100\,\AA\ is so small that it is hidden by the measurement 
uncertaintities.
The disappearance of small scale variations with increasing wavelength was 
also observed for NGC\,4151 in Dec.\,1993 
(cf.\,Edelson et al.\,\cite{Edelsonetal96}).

\subsection{Time-Series Analysis}
The emission lines are expected to change in response to variations in the 
far-UV continuum, primarily in response to variations at the unobservable 
wavelengths just shortward of 912\,\AA . Generally, it is necessary to assume 
that the observable continuum can approximate the behavior of the ionizing 
continuum and that the shortest observed UV wavelengths provide the best 
observable approximation to the ``driving'' (ionizing) continuum. 
However, for the first time we can make use of a well-sampled simultaneous 
X-ray light curve on short timescales (Chiang et al.\,\cite{Chiangetal00}). 
The EUVE light curve and the variations of the optical continuum show very 
similar variability pattern (Figs.\,3 and 4).

However, the H$\alpha$ and H$\beta$ emission line fluxes show only marginal
indications for variations. Within the errors the emission line flux was
constant within 5\%\ which is also indicated by F$_{var}$ (cf.\,Table 7).
In spite of this we calculated cross-correlation functions.
We used the extrem ultraviolet light curve measured with EUVE for the driving 
continuum kindly provided by Chiang and collaborators.
We computed the ICCF (cf.\,White \& Peterson \cite{WhPe94}) in the formalism
of the so called ``local CCF'' (Welsh \cite{Welsh99}). This approach takes 
into account the influence of the bias to underestimate the delay as has been 
shown by Welsh (\cite{Welsh99}). 
To correct for the influence of low frequency power in the flux variations 
we calculated a linear trend which was subtracted. The removal of a linear
trend helps to reduce the bias of the observed delay towards values smaller
than the real delay (Welsh \cite{Welsh99}). We also applied the 
cross-correlation analysis to the detrended light curves, and we calculated 
the discrete correlation function (DCF) for comparison. 
Uncertainties in the ICCF results for the cross-correlation maxima 
and centroids were computed through Monte Carlo techniques 
(cf.\,Peterson et al.\,\cite{BMPetal98}).

The sampling characteristics of each of the light curves are given in 
Table 8. The name of the feature and the total number of points, $N$, 
in the light curves are given that were used in computing the 
auto- and cross-correlation functions. The width (FWHM) of the ACF 
based on the original and of the detrended light curves are given as 
well. 
The corresponding widths of ICCF computed by cross-correlation with the 
EUV continuum at $\sim 0.2$ keV is listed too. 

\begin{figure}
\resizebox{\hsize}{!}{\includegraphics[bb=28 231 263 777,clip=true]{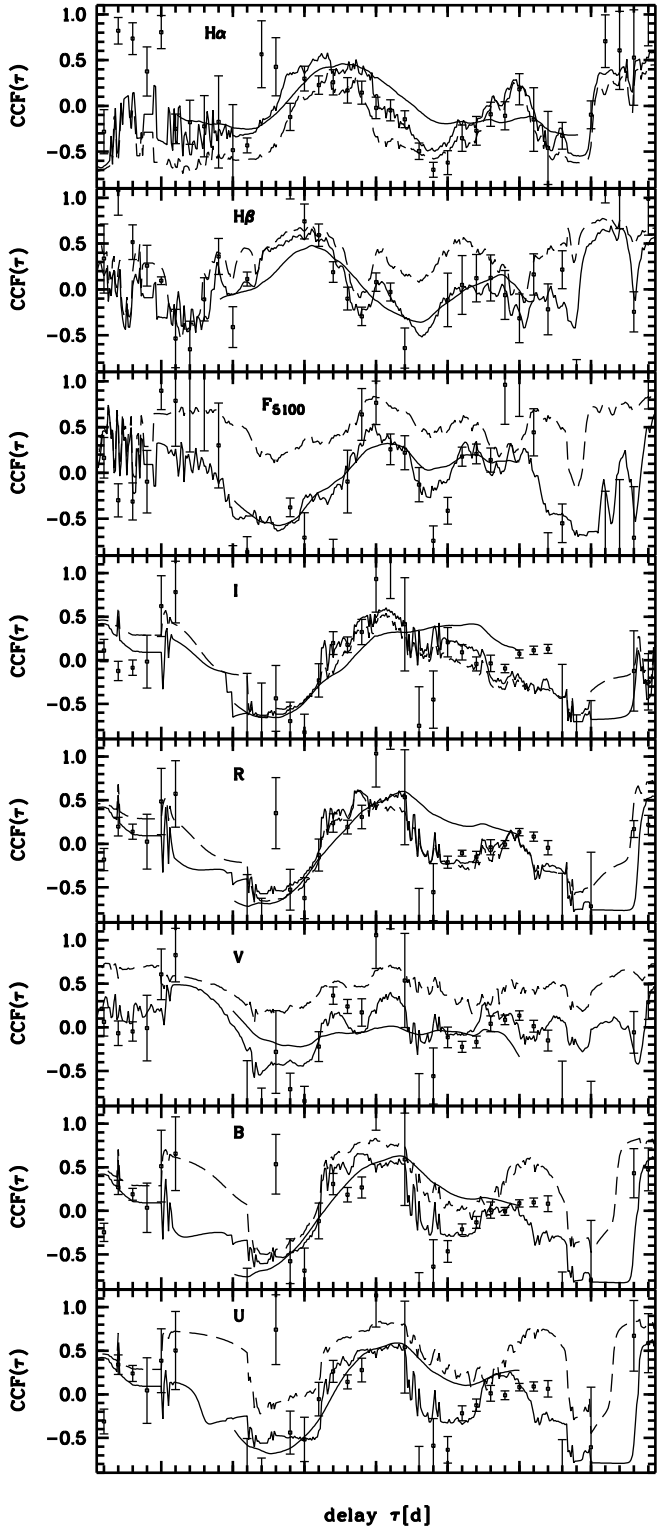}}
\caption{Cross-correlation functions for $U$,\,$B$,\,$V$,\,$R$,\,$I$
fluxes, F$_\lambda$(5100\,\AA), and the H$\beta $ and 
H$\alpha $ emission line (from bottom to top).
The ICCF computed with the original light curves (dashed line) and
the ICCF based on the detrended light curves (solid line) are displayed.
The average ICCFs which were obtained by estimating the uncertainties of the 
delays are shown as short solid line but restricted to -10$< \tau <$10\,d. 
The discrete points show the DCF computed for each detrended light curve.}
\end{figure}

\begin{table}
\caption[]{Sampling characteristics and full width at half maximum
           (FWHM) of ACF and CCF features. The FWHM is given in units 
           of days}
\begin{tabular}{lccccc}
\hline
\noalign{\smallskip}
feature   &N&\multicolumn{2}{c}{FWHM(ACF)}&\multicolumn{2}{c}{FWHM(ICCF)}\\
          & &orig.&detr.&orig.&detr.\\
\noalign{\smallskip}
\hline
$U$        &120&6.8&3.3&7.1&5.0\\
$B$        &120&5.7&4.2&6.6&5.9\\
$V$        &178&1.7&0.2&9.3&2.2\\
$R$        &120&1.8&1.6&4.1&5.9\\
$I$        &120&0.2&0.2&3.6&3.7\\
F$_\lambda$(5100\,\AA)& 52&30 &0.4&6.1&3.3\\
H$\beta$   & 52&1.7&0.6&4.0&2.4\\
H$\alpha$  & 38&1.2&0.2&1.1&1.4\\
\noalign{\smallskip}
\hline
\end{tabular}
\end{table}

The results of the cross-correlation analysis are given in Table 9.
The first column indicates the ``responding'' light curve (i.e., the light
curve that is assumed to be responding to the driving light curve). The
second and third column provide the peak value of the correlation coefficient 
$r_{\rm max}$ for the ICCF (original and detrended). The position of the 
peak of the cross-correlation functions $\tau _{\rm max}$ was measured by 
determing the location of the peak value of the ICCF and DCF; these values 
are given in the next four columns, respectively, 
The centroids $\tau _{\rm cent}$ of the ICCF of the original and detrended
light curves were computed using the points in the cross-correlation
function with values greater than $0.6r_{\rm max}$. 
The error estimate for the position of the cross-correlation peak 
$\Delta \tau_{\rm max}$ and the ICCF centroid $\Delta \tau_{\rm cent}$ 
for the original and detrended light curves are also given
(cf.\, Tab.\,9 columns $\Delta \tau _{orig.}$ and $\Delta \tau _{detr.}$).

The broad-band ($U$,\,$B$,\,$V$,\,$R$,\,$I$) and the optical continuum flux 
variations appear to be simultaneous within the errors, relative to the EUV 
continuum variations at $\sim 0.2$ keV (Fig.\,5). 
Taking into account the small variability amplitude expressed by 
F$_{\rm var}$ (Tab.\,7), the temporal coverage of the time series of 30
days only, with dense temporal sampling solely for the second half, 
the small CCF amplitudes, and the uncertainty of the location of the
peak and centroid of the CCFs of $\approx$3 days (Tab.\,9)
no reliable delay was detected for the broad emission lines of H$\alpha $ and 
H$\beta$ in response to the EUV variations (Fig.\,5).

\subsection{Mean and Root-Mean-Square Spectra}
We calculated mean and root-mean-square (rms) spectra from the flux-scaled
spectra for the H$\alpha$ and H$\beta$ regions.
The rms spectrum is useful for isolating variable parts of line profiles.\\
The mean and rms spectra of the samples C,\,L1,\,L2,\,W are identical within 
less than 3\% (H$\beta$) and 1.5\% (H$\alpha$). 
The mean and rms spectra of the H$\alpha$ and H$\beta$ region are presented 
for the spectra of subsample C (Fig.\,6). It provides 19 epochs during the 
studied period and a daily sampling with homogeneous settings for the second 
half of June 1998. 
The rms spectrum of H$\beta$ and H$\alpha$ indicates continuum variations of 
the order of less than 5\% . Furthermore, weak broad emission line flux 
variations are superimposed but the amplitude is less than 2\% . 
Thus, these features in the rms-spectra again should be taken only as 
weak indications for emission line flux variations on timescales of days. 
However, the width of the broad HeII\,4686 emission line in the rms spectrum 
is similar to that detected by Peterson et al.\,(\cite{BMPetal00}) for the 
narrow-line Seyfert\,1 galaxy NGC\,4051. The contribution of FeII emission 
(mulitplets 37,38) should be negligible since there is no indication of 

\begin{figure*}
\vspace*{9mm}
\resizebox{\hsize}{!}{\includegraphics{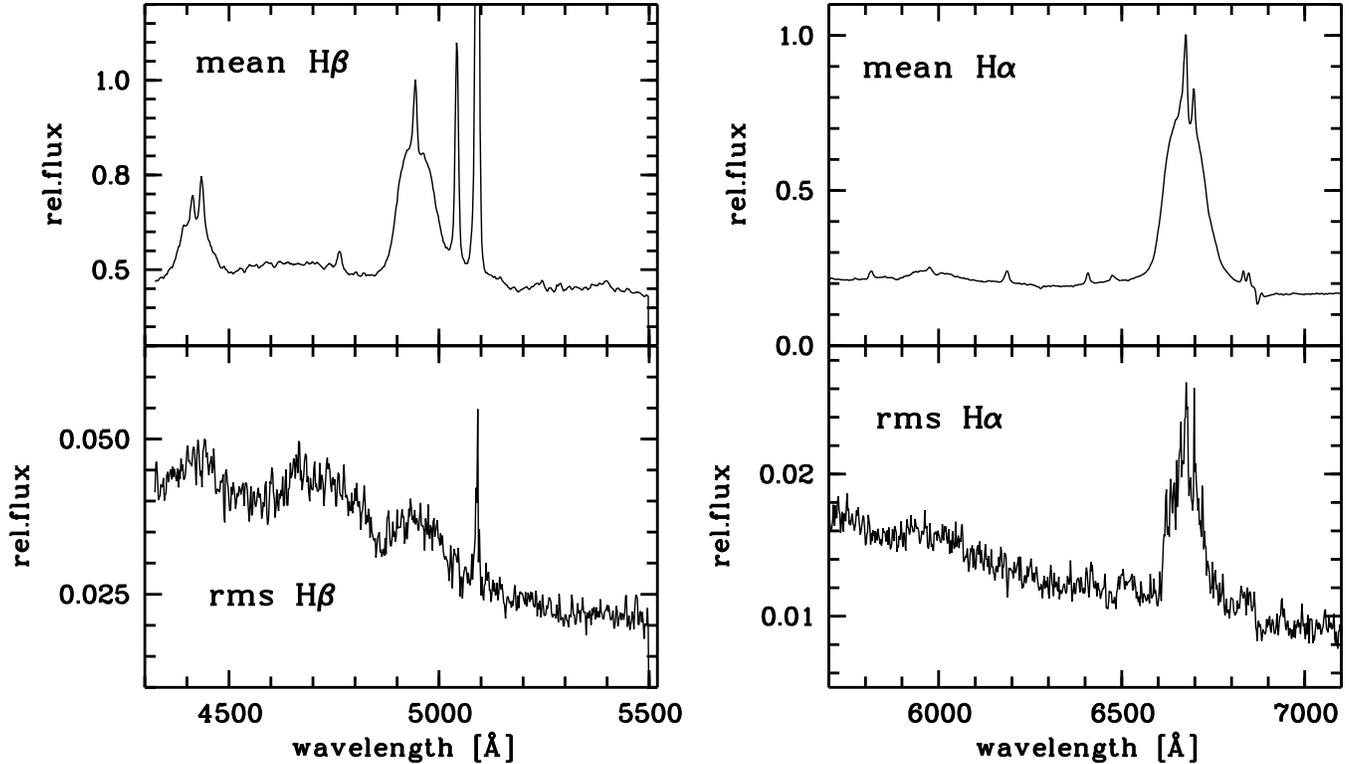}}
\caption{
The mean and root-mean-square (rms) spectra the of the 
H$\beta$ (top) and H$\alpha$ (bottom) line region are shown for sample C. 
The vertical scales are arbitrary. 
The rms spectrum of H$\beta$ and H$\alpha$ provide some indication for 
emission line variations but only of the order of less than 2\% . 
Remarkably the broad feature at the location of HeII\,4686 in the rms
spectrum of the H$\beta$ wavelength region. It is significantly broader than 
those of the H$\beta$ or H$\gamma$ line.}
\end{figure*}

\begin{table*}
\caption[]{Results of the cross-correlation analysis for the orginal and 
           detrended light curves.}
\begin{tabular}{lcccccccccc}
\hline
\noalign{\smallskip}
          &\multicolumn{2}{c}{r$_{max}$}
          &\multicolumn{4}{c}{$\tau _{max}$ [days]}
          &\multicolumn{2}{c}{$\tau _{cent}$ [days]}
          &\multicolumn{2}{c}{error.est.}\\
feature&orig.&detr.&\multicolumn{2}{c}{ICCF}&\multicolumn{2}{c}{DCF}&
    \multicolumn{2}{c}{ICCF}&\multicolumn{2}{c}{[days]}\\
&&&orig.&detr.&orig.&detr.&orig.&detr.&$\Delta \tau _{orig.}$&$\Delta \tau _{detr.}$\\
\noalign{\smallskip}
\hline
$U$       &0.86&0.57& 0.0& 0.9&0.0&1.0&-0.1&-0.1&0.6&1.1\\
$B$       &0.83&0.65&-0.1&-1.0&0.0&1.0&-0.4&-0.6&0.6&0.6\\
$V$       &0.71&0.40& 1.0& 1.0&-2.0&1.0&1.4& 1.0&2.1&2.1\\
$R$       &0.60&0.62&-1.2&-1.3&1.0&1.0&-0.0&-0.1&0.9&0.8\\
$I$       &0.52&0.58& 0.5& 0.5&1.0&1.0& 0.3& 0.6&5.2&2.0\\
$F_\lambda$(5100\,\AA)&0.83&0.53&-0.1&-0.1&0.0&0.0& 0.5& 0.1&1.4&3.2\\
H$\beta$  &0.56&0.04& 6.1& 6.5&6.0&7.0& 6.1& 7.8&3.0&4.5\\
H$\alpha$ &0.26&0.29& 9.8& 9.7&10.&10.& 9.8& 9.7&6.0&2.7\\
\noalign{\smallskip}
\hline
\end{tabular}
\end{table*}

\begin{figure}
\resizebox{\hsize}{!}{\includegraphics{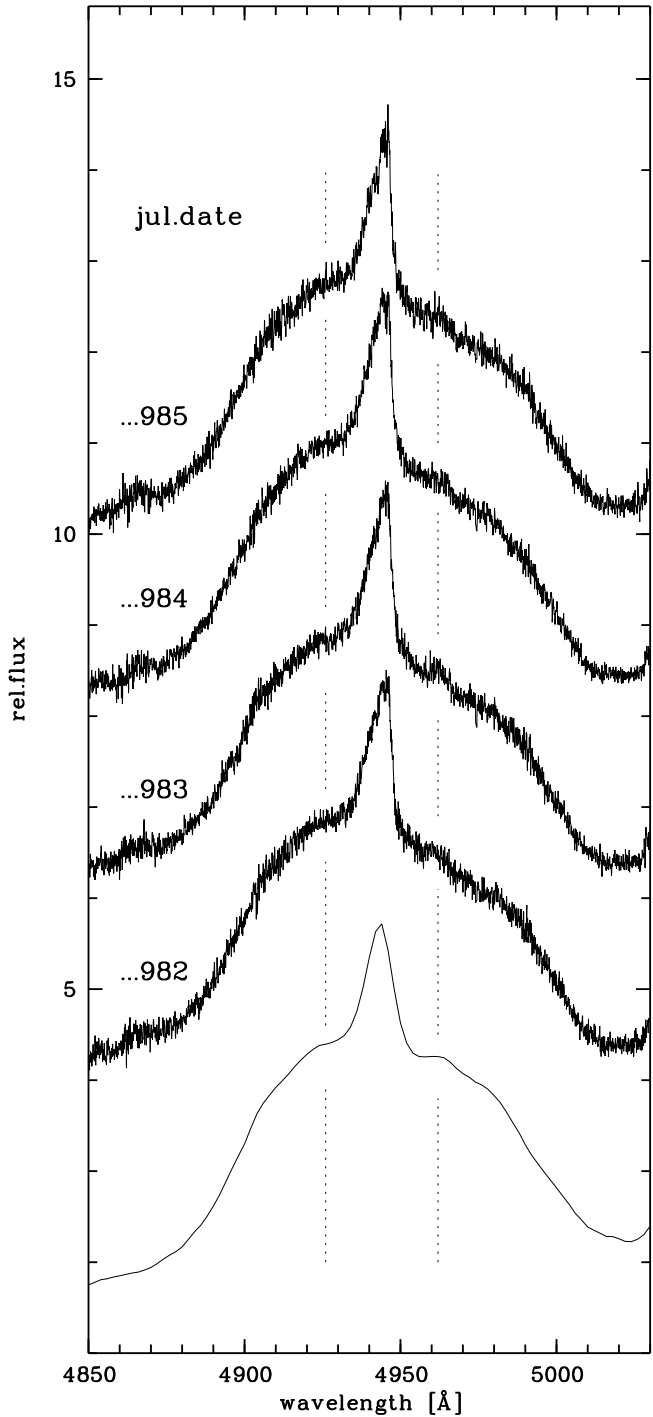}}
\caption{
The average echelle spectra of the H$\beta$ emission line profile as obtained 
at Calar Alto Observatory are shown for each night. For comparison the mean 
spectrum obtained from the samples C,\,L,\,W with lower spectal resolution 
is displayed at the bottom of the panel. The vertical dotted lines indicate 
the location of the marginal feature at $\lambda \simeq $4962\,\AA\ as well 
as the corresponding blueshifted location.}
\end{figure}

\noindent
variable FeII emission (multiplets 48,49) in the rms spectrum in the 
wavelength range $\lambda \lambda $ 5200\,--\,5500 \AA .

The spectra obtained at Calar Alto Observatory have a significant higher 
spectral resolution (R$\simeq $8\,km\,s$^{-1}$) than the spectra of the 
other samples (R$\simeq $500\,km\,s$^{-1}$).
Hence, we used the spectra of sample G to search for small features as 
suggested by Stella (\cite{Stella90}). The H$\beta$ profile was filtered 
with a narrow gaussian curve (FWHM=0.1\AA ) to improve the signal-to-noise 
ratio.
The average spectra of each night of the 15 epochs obtained at Calar Alto 
Observatory are presented in Fig.\,7 as well as the mean spectrum of the
samples C,\,L1,\,L2,\,W.
The shape of the H$\beta$ profile is smooth and no narrow features have been 
detected during this short period.
There is no obvious narrow moving feature visible in the H$\beta$ profile.
However, at $\lambda \simeq 4962$\AA\ there might be a marginal indication of
a weak structure which is visible in the mean spectrum and in most of the
individual spectra of sample G. It is redshifted by $\sim$1100\,km\,s$^{-1}$ 
and the width amounts to FWHM$\simeq $6\,\AA\ corresponding 
$\Delta v \simeq $350\,km\,s$^{-1}$.
We used also the spectra of the samples with lower spectral resolution to
search for this weak feature. It is also detectable in the mean spectrum 
based on these samples as can be seen in Fig.\,7.
Unfortunately, the redshift of v$\cong$1100\,km\,s$^{-1}$ places this 
weak feature in the red wing of the prominent [N\,{\sc II}]\,6583 
emission line. Hence, it is not possible to detect this structure in 
the H$\alpha$ line profile to provide further evidence for the existence 
of this weak feature.

\section{Summary}
The results of a short optical monitoring campaign (June 1998) 
on the Seyfert\,1 galaxy NGC\,5548 are presented in this paper. 
The principal findings are as follows:
\begin{enumerate}
\item The broad-band ($U$,$B$,$V$) fluxes, and the optical continuum 
      measured from spectrophotometry $F_{\lambda}$(5100\,\AA), showed a 
      monotonic decrease of approximately 35\,\% ($U$), 25\,\% ($B$), 
      15\,\% ($V$), and 30\,\% ($F_{\lambda}$(5100\,\AA)) 
      from the beginning to the end of the campaign ($\sim$ 30 days).
\item The  broad-band ($R$,$I$) fluxes, and the integrated emis\-sion-line 
      fluxes of H$\alpha$ and H$\beta$ showed only marginal indications for 
      variations on timescales of the order of a week with amplitudes less 
      than 5\% .
\item In all broad band fluxes a short flare like event was detected during 
      the night of June 22 (JD 2450987) which lasted $\approx$90 minutes.
\item The shape of the variability pattern of the optical continuum is very 
      similar to the detected variations in the EUV energy range 
      ($\sim 0.2$\,keV) with EUVE by Chiang et al.\,(\cite{Chiangetal00}). 
      But F$_{var}$ and R$_{\rm max}$ of the EUV variations are significantly 
      larger than F$_{var}$ and R$_{max}$ of F$_\lambda$(5100\,\AA ). 
\item The parameter $F_{\rm var}$, which is essentially the rms variation 
      about the mean, increased with decreasing wavelength for the
      broad-band measurements. The optical continuum $F_{\lambda}$(5100\,\AA)
      and the V- and R-band exhibits comparable $F_{\rm var}$.\\
      The Balmer emission lines H$\alpha$ and H$\beta$ underwent no 
      significant variations
      as measured by F$_{\rm var}$ during this campaign within the errors.
\item The variations of the broad-band and optical continuum fluxes 
      are simultaneous with the $\sim 0.2$\,keV variations. 
      In spite of the Balmer lines H$\alpha$ and H$\beta$ showing no 
      significant variability, we calculated cross-correlation functions.
      No reliable delay was detected for H$\alpha$ and H$\beta$ with respect 
      to the EUV continuum within the uncertainties. 
\item The H$\beta$ line profile was examined for narrow 
      features. There is marginal evidence for a weak feature 
      at $\lambda \simeq  $4962 \AA\ with FWHM$\simeq $6\,\AA\ which is
      redshifted by $\Delta $v$\simeq $1100\,km\,s$^{-1}$ with respect to 
      H$\beta _{\rm narrow}$.
\end{enumerate}

\begin{acknowledgements} 
This work has been supported by SFB328D (Landessternwarte Heidelberg), 
by the NASA grant NAG\,5-3234 (University of Florida), 
the  Russian Basic Research Foundation grant N 94-02-4885a, N 97-02-17625
(Sternberg Astronomical Institute, Special Astrophysical Observatory),
by the Smithsonian Institution, and by INTAS grant N96-032.
We would like to thank the FLWO remote observers P.\,Berlind and 
M.\,Calkins, and also S.\,Tokarz for help in reducing and archiving
the FLWO data.
\end{acknowledgements}

\end{document}